\documentclass[proceedings]{stacs}
\stacsheading{2009}{255--264}{Freiburg}
\firstpageno{255}

\usepackage{amsmath, amsthm}
\usepackage{amsfonts, enumerate}

\newenvironment{cpitemize}
{\ifvmode\vspace{-2ex}\fi\begin{itemize}\setlength{\itemsep}{-0.5ex}}
  {\end{itemize}\vspace{-1ex}}

\newcommand{\ceil}[1]{\lceil #1 \rceil}

\newcommand{\SETF}{{\rm SETF}}
\newcommand{\LAPS}{{\rm LAPS}}
\newcommand{\SRPT}{{\rm SRPT}}

\newcommand{\RR}{{\rm RR}}
\newcommand{\EA}{\text{$\LAPS(\delta, \beta)$}}
\newcommand{\Xomit}[1]{ }

\title{Nonclairvoyant Speed Scaling for Flow and Energy}

\author[lab1]{H.L. Chan}{Ho-Leung Chan}
\address[lab1]{Max-Planck-Institut f\"ur Informatik}  
\email[H.L. Chan]{hlchan@mpi-inf.mpg.de}

\author[lab4]{J.Edmonds}{Jeff Edmonds} 
\address[lab4]{Department of Computer Science and Engineering,
York University}
\email[J.Edmonds]{jeff@cse.yorku.ca}

\author[lab2]{T.W.Lam}{Tak-Wah Lam}
\address[lab2]{Department of Computer Science,
University of Hong Kong}	
\email[T.W.Lam]{twlam@cs.hku.hk}
\email[L.K. Lee]{lklee@cs.hku.hk}  

\author[lab2]{L.K. Lee}{Lap-Kei Lee} 

\author[lab3]{A.Marchetti-Spaccamela}{Alberto Marchetti-Spaccamela}
\address[lab3]{Dipartimento di Informatica e Sistemistica,
Sapienza Universit\`a di Roma}  
\email[A.Marchetti-Spaccamela]{alberto@dis.uniroma1.it} 

\author[lab5]{K.Pruhs}{Kirk Pruhs}
\address[lab5]{Computer Science Department,
University of Pittsburgh}
\email[K.Pruhs]{kirk@cs.pitt.edu}

\thanks{The work of H.L.Chan was done when he was a
postdoc in University of Pittsburgh. 
T.W.Lam is  partially supported by HKU Grant 7176104. 
A.Marchetti-Spaccamela is partially  supported by MIUR FIRB grant  RBIN047MH9   and by EU ICT-FET grant 215270 FRONTS. 
K.Pruhs is partially  supported  by  an IBM faculty award, and 
from NSF grants CNS-0325353, CCF-0514058, IIS-0534531, and CCF-0830558.}

\begin{document}
\date{}

\begin{abstract}
We study online nonclairvoyant speed scaling to minimize
total flow time plus energy. We first consider the traditional
model where the power function is $P(s)=s^\alpha$.
We give a nonclairvoyant algorithm that is shown to be $O( \alpha^3 )$-competitive.
We then show an
$\Omega( \alpha^{1/3-\epsilon} )$ lower bound on the competitive
ratio of any nonclairvoyant algorithm.
We also show that there
are power functions for which no nonclairvoyant algorithm can be $O(1)$-competitive.
\end{abstract}
\maketitle

\section{Introduction}

Energy consumption has become a key issue in the design of microprocessors.
Major chip manufacturers, such as Intel, AMD and IBM,
now produce chips with dynamically scalable speeds,
and produce associated software, such as Intel's SpeedStep and AMD's PowerNow,
that enables an operating system to manage
power by scaling processor speed. Thus the operating system should
have an {\em speed scaling} policy for setting the speed
of the processor, that ideally should work in tandem with a
{\em job selection} policy for determining which job to run.
The operating system has dual competing objectives, as it both
wants to optimize some schedule quality of service objective,
as well as some power related objective.

In this paper, we will consider the objective of minimizing a linear combination of 
total flow and
total energy used. 
For a formal definitions of the problem that we consider,
see subsection \ref{subsec:formaldefinitions}. 
This objective of flow plus energy
has a natural interpretation. Suppose that the user specifies how much improvement in flow,
call this amount $\rho$, is necessary
to justify spending one unit of energy. 
For example, the user might specify that he is willing
to spend 1 erg of energy from the battery 
for a decrease of 5 micro-seconds in flow.
Then the optimal schedule, from this user's perspective, is the schedule
that optimizes  $\rho=5$ times the energy used plus the total flow.
By changing the units of either energy or time, 
one may assume without loss of generality that $\rho=1$.

In order to be implementable in a real system, the speed scaling and job selection
policies must be online since the
system will not in general know about jobs arriving in the future.
Further, to be implementable in a generic operating system, these policies must
be nonclairvoyant, since in general the operating system does not know the
size/work of each process when the process is released to the operating system. 
All of the previous speed scaling literature on
this objective has considered either offline or online clairvoyant policies. 
In subsection \ref{subsec:survey}, we
survey the literature on nonclairvoyant scheduling policies for flow objectives 
on fixed speed processors, and the speed scaling literature for flow plus energy objectives.

Our goal in this paper is to study nonclairvoyant speed scaling assuming an off-line adversary 
that dynamically chooses the speed of its own machine.  

We first analyze the nonclairvoyant algorithm whose job selection policy is Latest Arrival Processsor 
Sharing ($\LAPS$) and whose speed scaling policy is to run at speed
$(1+\delta)$ times the number of active
jobs. $\LAPS$ shares the processor equally among the latest arriving constant fraction
of the jobs. 
We adopt the traditional model that the power function, which
gives the power as a function of the speed of the processor,
is $P(s)=s^\alpha$,
where $\alpha > 1$ is some constant. Of particular interest is the case that
$\alpha=3$ since according to the well known cube-root rule, the dynamic power in CMOS based processors is
approximately the cube of the speed. 
Using an amortized local competitiveness argument, we show in section \ref{sec:upper} that
this algorithm is $O( \alpha^3 )$-competitive. 
The potential function that we use is an amalgamation of the potential function
used in \cite{EP} for the fixed speed analysis of $\LAPS$, and the potential functions
used for analyzing clairvoyant speed scaling policies. 
This result shows that it is possible for a nonclairvoyant policy to be
$O(1)$-competitive if the cube-root rule holds.

It is known  that for essentially every power function,
there is a 3-competitive {\em clairvoyant} speed scaling policy \cite{BCPSODA2009}.
In contrast, we show that the competitiveness achievable by
{\em nonclairvoyant} policies must depend on the power function. 
In the traditional model, we show in section \ref{sec:lower} 
an $\Omega( \alpha^{1/3-\epsilon} )$ lower bound on the competitive ratio
of any deterministic nonclairvoyant algorithm.
Further, we show in section \ref{sec:lower} that there exists a particular power function for which
there is no $O(1)$-competitive deterministic nonclairvoyant speed scaling
algorithm. The adversarial strategies for these lower bounds 
are based on the adversarial strategies
in \cite{MPT} for fixed speed processors.
Perhaps these lower bound results are not so surprising given the fact
that it is known that without speed scaling, resource augmentation is required to achieve 
$O(1)$-competitiveness for a nonclairvoyant policy~\cite{MPT,KP:Speedy}.
Still a priori it wasn't completely clear that the lower bounds in \cite{MPT}
would carry over. The reason is that in these lower bound instances, the
adversary forced the online algorithm into a situation in which the online algorithm had
a lot of jobs with a small amount of remaining work, while the adversary had
one job left with a lot of remaining work. In the fixed speed setting, the
online algorithm, without resource augmentation, can never get a chance to 
get rid of this backlog in the face of a steady stream of jobs.
However, in a speed scaling setting, one might imagine 
an online algorithm that speeds up enough to remove the backlog, but not enough to 
make its energy usage more than a constant time optimal. Our lower
bound shows that it is not possible for the online algorithm to accomplish this.


\subsection{Related results}
\label{subsec:survey}

We start with some results in the literature about 
scheduling with the objective of total flow time 
on a fixed speed processor.
It is well known that the online clairvoyant algorithm Shortest Remaining Processing Time
($\SRPT$) is optimal. The competitive ratio of deterministic nonclairvoyant algorithm
is $\Omega(n^{1/3})$, and the competitive ratio of every randomized algorithm against
an oblivious adversary is $\Omega(\log n)$~\cite{MPT}. A randomized version of 
the Multi-Level Feedback Queue algorithm is $O(\log n)$-competitive \cite{KPRMLF,BLRMLF}.
The non-clairvoyant algorithm Shortest Elapsed Time First ($\SETF$) is scalable, that is,
$(1+\epsilon)$-speed $O(1)$-competitive~\cite{KP:Speedy}.
$\SETF$ shares the processor equally among all jobs that have been run the least.
The algorithm Round Robin $\RR$ (also called Equipartition and Processor Sharing) 
that  shares the processor equally among all jobs is 
$(2+\epsilon)$-speed $O(1)$-competitive~\cite{EdEqui}.

Let us first consider the traditional model where the
power function is $P=s^\alpha$. 
Most of the literature assumes the {\em unbounded speed model}, in which a processor
 can be run at any real speed in the range $[0, \infty)$. 
So let us now consider the unbounded speed model.
\cite{PUW} gave an efficient offline algorithm to find the schedule that
minimizes average flow subject to a constraint on the amount of
energy used, in the case that jobs have unit work. This algorithm
can also be used to find optimal schedules when the objective
is a linear combination of total flow and energy used.
\cite{PUW} observed that in any locally-optimal schedule,
essentially each job $i$ is run at a power proportional to the number of
jobs that would be delayed if job $i$ was delayed.  
\cite{AF} proposed the natural online speed scaling
algorithm that always runs at a power equal to the number of unfinished
jobs (which is lower bound to the number of jobs that would be delayed if the
selected job was delayed). 
\cite{AF} did not actually analyze this natural algorithm, but rather analyzed
a batched variation, in which jobs that are released while the current
batch is running are ignored until the current batch finishes.
\cite{AF} showed that for unit work jobs this batched algorithm
is $O\left(\left(\frac{3 + \sqrt{5}}{2}\right)^\alpha\right)$-competitive
by reasoning directly
about the optimal schedule. 
\cite{AF} also gave an efficient offline dynamic programming algorithm.
\cite{BPS} considered the  algorithm that runs at a power 
equal to the unfinished work
(which is in general a bit less than the number of unfinished jobs for
unit work jobs).
\cite{BPS} showed that for unit work jobs, 
this algorithm is 2-competitive with respect to the
objective of fractional flow plus energy using 
an amortized local competitiveness argument.
\cite{BPS} then showed that the natural algorithm proposed in \cite{AF}
is 4-competitive for total flow plus energy for unit work jobs.

In \cite{BPS} the more general setting where jobs have arbitrary sizes and arbitrary weights and the objective is weighted flow plus energy has been considered.
The authors  analysed the algorithm that uses
Highest Density First (HDF) for job selection, and always runs at a power equal to the
fractional weight of the unfinished jobs.
\cite{BPS} showed that this algorithm is $O(\frac{\alpha}{\log \alpha})$-competitive for fractional weighted
flow plus energy using an amortized local competitiveness argument.
\cite{BPS} then showed how to modify this
algorithm to obtain an algorithm that is $O(\frac{\alpha^2}{\log^2 \alpha})$-competitive for (integral) weighted flow
plus energy 
using the known resource augmentation analysis of HDF~\cite{BLMP1}.

Recently, \cite{LLTW08} improves on the obtainable competitive ratio for total flow plus energy for arbitrary work
and unit weight jobs by  considering the job selection algorithm Shortest Remaining
Processing Time (SRPT) and the speed scaling algorithm of running at a power proportional to the
number of unfinished jobs. 
\cite{LLTW08} proved that this algorithm is
$O(\frac{\alpha}{\log \alpha})$-competitive for arbitrary size and unit
weight jobs.

In \cite{BCLL08} the authors extended the results of \cite{BPS} for the unbounded speed model to
the bounded speed model, where there is an upper bound on the processor speed. 
The speed scaling algorithm was to run at the minimum
of the speed recommended by the speed scaling algorithm in the unbounded speed
model and the maximum speed of the processor. 
The results for the bounded speed model in \cite{BCLL08} were improved in \cite{LLTW08} proving  
 competitive ratios
of the form $O(\frac{\alpha}{\log \alpha})$.

\cite{BCPSODA2009} consider a more general model.
They assume that the allowable speeds are a countable collection
of disjoint subintervals of $[0, \infty)$,
and consider arbitrary power functions $P$ that are non-negative, and 
continuous and differentiable on all but countably many points.
They give two main results in this general model.
The scheduling algorithm, that uses Shortest Remaining Processing Time (SRPT) 
for job selection and power equal to one more than the number
of unfinished jobs for speed scaling, 
is $(3+\epsilon)$-competitive for the objective of
total flow plus energy on arbitrary-work unit-weight jobs.  
The scheduling algorithm, that uses Highest Density First (HDF) for job selection and power equal to the fractional
weight of the unfinished jobs for speed scaling,
is 
$(2+\epsilon)$-competitive for the objective of
fractional weighted flow plus energy on arbitrary-work arbitrary-weight jobs.

\subsection{Formal Problem Definition and Notations}
\label{subsec:formaldefinitions}
We study online scheduling on a single processor.
Jobs arrive over time and we have no information
about a job until it arrives.
For each job $j$, its release time and work requirement
(or size) are denoted as $r(j)$ and $p(j)$, respectively.
We consider the {\em nonclairvoyant} model, i.e.,
when a job $j$ arrives, $p(j)$ is not given and
it is known only when $j$ is completed.
Preemption is allowed and has no cost;
a preempted job can resume at the point of preemption.
The processor can vary its speed dynamically to any value in $[0, \infty)$.
When running at speed $s$, the processor processes $s$ units of work per unit time
and consumes $P(s) = s^\alpha$ units of energy per unit time, where $\alpha > 1$
is some fixed constant. We call $P(s)$ the \emph{power function}.

Consider any job sequence $I$ and a certain schedule $A$ of $I$.
For any job $j$ in $I$, the flow time of $j$, denoted $F_A(j)$,
is the amount of time elapsed since it arrives until it is completed.
The total flow time of the schedule is $F_A = \sum_{j\in I} F_A(j)$.
We can also interpret $F_A$ as follows. Let $n_A(t)$
be the number of jobs released by time $t$ but not yet completed by time $t$.
Then $F_A = \int_{0}^{\infty} n_A(t) dt$.
Let $s_A(t)$ be the speed of the processor at time $t$ in the schedule.
Then the total energy usage of the schedule is $E_A = \int_{0}^{\infty} (s(t))^\alpha dt$.
The objective is to minimize the sum of total flow time and energy usage,
i.e., $F_A + E_A$.

For any job sequence $I$,
a scheduling algorithm ALG needs to specify
at any time the speed of the processor and the jobs
being processed.
We denote ALG$(I)$
as the schedule produced for $I$ by ALG.
Let Opt be the optimal offline algorithm such that for any job sequence $I$,
$F_{Opt(I)} + E_{Opt(I)}$ is minimized among all schedules of $I$.
An algorithm ALG is said to be $c$-competitive, for any $c \ge 1$,
if for all job sequence $I$,
\[
F_{ALG(I)} + E_{ALG(I)} \le c \cdot ( F_{Opt(I)} + E_{Opt(I)} )
\]

\section{An $O( \alpha^3)$-competitive Algorithm}
\label{sec:upper}
In this section,
we give an online nonclairvoyant algorithm
that is $O(\alpha^3)$-competitive for total flow time plus energy.
We say a job $j$ is \emph{active} at time
$t$ if $j$ is released by time $t$ but not yet completed
by time $t$.  Our algorithm is defined as follows.
\begin{quote}
{\bf Algorithm $\EA$.}
Let $0 < \delta, \beta \le 1$ be any real.
At any time $t$, the processor speed is $(1+ \delta ) (n(t))^{1/\alpha}$,
where $n(t)$ is the number of active jobs at time $t$.
The processor processes the $\ceil{\beta n(t)}$ active jobs with the latest release times
 (ties are broken by job ids) by splitting the processing speed equally among these jobs.
\end{quote}
Our main result is the following.

\begin{theorem}\label{thm:ea}
When $\delta = \frac{3}{\alpha}$ and $\beta = \frac{1}{2\alpha}$,
$\EA$ is $c$-competitive for total flow time plus energy,
where $c = 4\alpha^3(1+(1+\frac{3}{\alpha})^\alpha) = O(\alpha^3)$.
\end{theorem}

The rest of this section is devoted to proving Theorem~\ref{thm:ea}.
We use an amortized local competitiveness argument (see for example \cite{Pruhs07}).
To show that an algorithm is $c$-competitive  it is sufficient to show  a potential function such that at any time $t$ 
the increase in the objective cost of the algorithm  plus the change of the potential is 
at most $c$ times the increase in the objective of the optimum.

For any time $t$,
let $G_a(t)$ and $G_o(t)$ be the total flow time plus energy
incurred up to time $t$ by $\EA$
and the optimal algorithm Opt, respectively.
To show that $\EA$ is $c$-competitive,
it suffices to give a potential function $\Phi(t)$ such that
the following four conditions hold.
\begin{cpitemize}
\item
{\em Boundary condition:}
$\Phi = 0$ before any job is released and $\Phi \ge 0$ after
all jobs are completed.

\item
{\em Job arrival:}
When a job is released,
$\Phi$ does not increase.

\item
{\em Job completion:}
When a job is completed by $\EA$ or OPT,
$\Phi$ does not increase.

\item
{\em Running condition:}
At any other time, the rate of change of $G_a$ plus that of $\Phi$
is no more than $c$ times the rate of change of $G_o$.
That is,
$\frac{d G_a(t)}{dt} + \frac{d \Phi(t)}{dt} \le c\cdot \frac{d G_o(t)}{dt}$
during any period of time without job arrival or completion.
\end{cpitemize}
Let $n_a(t)$ and $s_a(t)$ be the number of active jobs
and the speed in $\EA$ at time $t$, respectively.
Define $n_o(t)$ and $s_o(t)$ similarly for that of Opt.
Then 
$$\frac{d G_a(t)}{dt} = \frac{d F_{LAPS}(t)}{dt}+ E_{LAPS}(t) =  n_a(t) + (s_a(t))^\alpha$$
and, similarly, $\frac{d G_o(t)}{dt} = n_o(t) + (s_o(t))^\alpha$.
We define our potential function as follows.

\begin{quote}
{\bf Potential function $\Phi(t)$.}
Consider any time $t$.
For any job $j$, let $q_a(j,t)$ and $q_o(j, t)$
be the remaining work of $j$ at time $t$ in $\EA$ and Opt, respectively.
Let $\{j_1, \ldots, j_{n_a(t)}\}$ be the set of active jobs in $\EA$,
ordered by their release time such that
$r(j_1) \le r(j_2) \le \dots \le r(j_{n_a(t)})$.
Then,
\[
 \Phi(t) = \gamma \sum_{i=1}^{n_a(t)}
 \left( i^{1-1/\alpha} \cdot \max\{0, q_a(j_i, t) - q_o(j_i, t)\} \right)
\]
where $\gamma = \alpha(1+(1+\frac{3}{\alpha})^\alpha)$.
We call $i^{1-1/\alpha}$ the {\em coefficient} of $j_i$.
\end{quote}

We first check the boundary, job arrival and job completion
conditions.
Before any job is released or after all jobs are completed,
there is no active job in both $\EA$ and Opt,
so $\Phi = 0$ and the boundary condition holds.
When a new job $j$ arrives at time $t$, $q_a(j, t) - q_o(j, t) = 0$ and the coefficients
of all other jobs remain the same, so $\Phi$ does not change.
If $\EA$ completes a job $j$,
the term for $j$ in $\Phi$ is removed.
The coefficient of any other job
either stays the same or decreases, so $\Phi$ does not increase.
If Opt completes a job, $\Phi$ does not change.

It remains to check the running condition. In the following,
we focus on a certain time $t$ within a period of time
without job arrival or completion.
We omit the parameter $t$ from the notations
as $t$ refers only to this certain time.
For example, we denote $n_a(t)$ and $q_a(j, t)$ as
$n_a$ and $q_a(j)$, respectively.
For any job $j$, if $\EA$ has processed less than Opt on $j$ at time $t$,
i.e., $q_a(j) - q_o(j) > 0$,
then we say that $j$ is a \emph{lagging job} at time $t$.
We start by evaluating $\frac{d\Phi}{dt}$.

\begin{lemma}\label{lem:phi}
Assume $\delta = \frac{3}{\alpha}$ and $\beta = \frac{1}{2\alpha}$.
At time $t$,
if $\EA$ is processing less than $(1 - \frac{1}{2\alpha}) \ceil{\beta n_a}$ lagging jobs,
then $\frac{d\Phi}{dt} \le \frac{\gamma}{\alpha} s_o^\alpha + \gamma (1-\frac{1}{\alpha}) n_a$.
Else if $\EA$ is processing at least $(1 - \frac{1}{2\alpha}) \ceil{\beta n_a}$ lagging jobs,
then $\frac{d\Phi}{dt} \le \frac{\gamma}{\alpha} s_o^\alpha - \frac{\gamma}{\alpha} n_a$.
\end{lemma}

\begin{proof}
We consider $\frac{d\Phi}{dt}$ as the combined effect due
to the processing of $\EA$ and Opt.
Note that for any job $j$, $q_a(j)$ is decreasing at a rate of either
0 or $- s_a / \ceil{\beta n_a}$.
Thus the rate of change of $\Phi$ due to $\EA$ is non-positive.
Similarly, the rate of change of $\Phi$ due to Opt is non-negative.

We first bound the rate of change of $\Phi$ due to Opt.
The worst case is that Opt is processing the job with the largest coefficient,
i.e., $n_a^{1-1/\alpha}$.
Thus the rate of change of $\Phi$ due to Opt is at most
$\gamma n_a^{1-1/\alpha} (-\frac{d q_o(j_{n_a})}{dt}) = \gamma n_a^{1-1/\alpha} s_o$.
We apply Young's Inequality \cite{HLP52}, which is formally stated
in Lemma~\ref{lem:young},
by setting $f(x)=x^{\alpha-1}$, $f^{-1}(x)=x^{1/(\alpha -1)}$,
$g=s_o$ and $h=n_a^{1-1/\alpha}$. Then, we have
\[
s_o n_a^{1-1/\alpha} \le
\int_0^{s_o} x^{\alpha-1} dx + \int_0^{n_a^{1-1/\alpha}} x^{1/(\alpha -1)} dx =
\frac{1}{\alpha} s_o^\alpha+ (1 - \frac{1}{\alpha}) n_a
\]

If $\EA$ is processing less than $(1 - \frac{1}{2\alpha}) \ceil{\beta n_a}$ lagging jobs,
we just ignore the effect due to $\EA$ and take the bound
that $\frac{d\Phi}{dt} \le \frac{\gamma}{\alpha} s_o^\alpha + \gamma (1 - \frac{1}{\alpha}) n_a$.

If $\EA$ is processing at least $(1 - \frac{1}{2\alpha}) \ceil{\beta n_a}$ lagging jobs,
let $j_i$ be one of these lagging jobs.
We notice that $j_i$ is among the $\ceil{\beta n_a}$ active jobs with
the latest release times.
Thus, the coefficient of $j_i$ is at least $(n_a - \ceil{\beta n_a} + 1)^{1-1/\alpha}$.
Also, $j_i$ is being processed at a speed of $s_a/\ceil{\beta n_a}$,
so $q_a( j_i, t)$ is decreasing at this rate.
$\EA$ is processing at least $(1 - \frac{1}{2\alpha}) \ceil{\beta n_a}$ such lagging jobs,
so the rate of change of $\Phi$ due to $\EA$ is more negative than
\begin{eqnarray*}
&&\gamma \left((1 - \frac{1}{2\alpha})\ceil{\beta n_a}\right)
  \left(n_a - \ceil{\beta n_a} + 1\right)^{1 - 1/\alpha}
  \left(\frac{-s_a}{\ceil{\beta n_a}}\right) \\
&\le& - \gamma (1 - \frac{1}{2\alpha}) (n_a - \beta n_a)^{1 - 1/\alpha} (s_a)
\hspace{.88in}\mbox{(since $-\ceil{\beta n_a} + 1 \ge -\beta n_a$)}\\
&\le& - \gamma  (1 - \frac{1}{2\alpha}) ( 1- \beta) ( 1+ \delta) n_a
\hspace{1.1in}\mbox{(since $s_a = (1+\delta) n_a^{1/\alpha}$)}
\end{eqnarray*}
When $\beta = \frac{1}{2\alpha}$ and $\delta = \frac{3}{\alpha}$,
simple calculation shows that
$(1 - \frac{1}{2\alpha}) ( 1- \beta) ( 1+ \delta) \ge 1$ and
hence the last term above is at most $-\gamma n_a$.
It follows that
$\frac{d\Phi}{dt}
\le \frac{\gamma}{\alpha} s_o^\alpha + \gamma (1 - \frac{1}{\alpha}) n_a
- \gamma n_a
= \frac{\gamma}{\alpha} s_o^\alpha - \frac{\gamma}{\alpha} n_a$.
\end{proof}

Below is the formal statement of Young's Inequality, which is used in the proof of Lemma~\ref{lem:phi}.

\begin{lemma}[Young's Inequality \cite{HLP52}]\label{lem:young}
Let $f$ be any real-value, continuous and strictly increasing function $f$ such that $f(0)=0$.
Then, for all $g, h \ge 0$,
$\int_0^g f(x) dx + \int_0^h f^{-1}(x) dx \ge gh$,
where $f^{-1}$ is the inverse function of $f$.
\end{lemma}

We are now ready to show the following lemma about the running condition.
\begin{lemma}\label{lem:running}
Assume $\delta = \frac{3}{\alpha}$ and $\beta = \frac{1}{2\alpha}$.
At time $t$, $\frac{d G_a}{dt} + \frac{d \Phi}{dt} \le c \cdot \frac{d G_o}{dt}$,
where $c = 4\alpha^3(1+(1+\frac{3}{\alpha})^\alpha)$.
\end{lemma}

\begin{proof}
We consider two cases depending on the number of lagging jobs that
$\EA$ is processing at time $t$.
If $\EA$ is processing at least $(1 - \frac{1}{\alpha}) \ceil{\beta n_a}$ lagging jobs,
then 
\begin{eqnarray*}
\frac{d G_a}{dt} + \frac{d \Phi}{dt} &=&
n_a + s_a^\alpha + \frac{d\Phi}{dt} \\
&\le& n_a +(1+\delta)^\alpha n_a + \frac{\gamma}{\alpha} s_o^\alpha - \frac{\gamma}{\alpha} n_a \hspace{1in}\mbox{(by Lemma~\ref{lem:phi})}\\
& = & (1 + (1+\delta)^\alpha - \frac{\gamma}{\alpha}) n_a + \frac{\gamma}{\alpha} s_o^\alpha
\end{eqnarray*}
Since $\delta = \frac{3}{\alpha}$ and $\gamma = \alpha( 1 + (1+\frac{3}{\alpha})^\alpha )$, the coefficient of $n_a$ becomes zero
and $\frac{d G_a}{dt} + \frac{d \Phi}{dt} \le \frac{\gamma}{\alpha} s_o$.
Note that $\frac{\gamma}{\alpha}
= (1 + (1+\frac{3}{\alpha})^\alpha ) \le c$ and $\frac{dG_o}{dt} = n_o + s_o^\alpha$,
so we have $\frac{d G_a}{dt} + \frac{d \Phi}{dt} \le c \cdot \frac{dG_o}{dt}$.

If $\EA$ is processing less than $(1 - \frac{1}{2\alpha}) \ceil{\beta n_a}$ lagging jobs,
the number of jobs remaining in Opt is $n_o \ge \ceil{\beta n_a} - (1 - \frac{1}{2\alpha}) \ceil{\beta n_a}
= \frac{1}{ 2 \alpha} \ceil{\beta n_a} \ge \frac{1}{2\alpha} \beta n_a
= \frac{1}{4\alpha^2} n_a $.
Therefore, 
\begin{eqnarray*}
\frac{d G_a}{dt} + \frac{d \Phi}{dt} &=&
n_a + s_a^\alpha + \frac{d\Phi}{dt} \\
&\le& n_a + (1+\delta)^\alpha n_a + \frac{\gamma}{\alpha} s_o^\alpha + \gamma (1 - \frac{1}{\alpha}) n_a \hspace{1in}\mbox{(by Lemma~\ref{lem:phi})}\\
& =& (1 + (1+\delta)^\alpha + \gamma(1-\frac{1}{\alpha})) n_a
+ \frac{\gamma}{\alpha}s_o^\alpha \\
&\le& 4 \alpha^2(1 + (1+\delta)^\alpha + \gamma(1-\frac{1}{\alpha})) n_o
+ \frac{\gamma}{\alpha}s_o^\alpha
\end{eqnarray*}
Since $\delta = \frac{3}{\alpha}$ and $\gamma = \alpha( 1 + (1+\frac{3}{\alpha})^\alpha)$,
the coefficient of $n_o$ becomes $4\alpha^3( 1 + ( 1+\frac{3}{\alpha})^\alpha) = c$.
The coefficient of $s_o^\alpha$ is $( 1 + ( 1+\frac{3}{\alpha})^\alpha) \le c$.
Since $\frac{dG_o(t)}{dt} = n_o + s_o^\alpha$,
we obtain $\frac{d G_a(t)}{dt} + \frac{d \Phi}{dt} \le c \cdot \frac{dG_o(t)}{dt}$.
Note that this case is the bottleneck leading to the current competitive ratio.
\end{proof}

Combining Lemma~\ref{lem:running} with the discussion on
the boundary, job arrival and job completion conditions,
Theorem~\ref{thm:ea} follows.

\section{Lower Bounds}
\label{sec:lower}
In this section, we show that every nonclairvoyant algorithm is $\Omega( \alpha^{1/3 -\epsilon})$-competitive
in the traditional model where
the power function
$P(s) = s^\alpha$.
We further extend the lower bound to other power functions $P$
and show that for some power function, any algorithm
is $\omega(1)$-competitive. We first prove the following lemma.

\begin{lemma}\label{lem:lower}
Let $P(s)$ be any non-negative, continuous and super-linear power function.
Let $k, v \ge 1$ be any real such that $P(v) \ge 1$.
Then, any algorithm is $\Omega( \min\{k, P( v + \frac{1}{16(k P(v))^3})/P(v) \})$-competitive.
\end{lemma}

\begin{proof}
Let ALG be any algorithm and Opt be the offline adversary.
Let $n = \ceil{ k P(v)}$.
We release $n$ jobs $j_1, j_2, \dots, j_n$ at time 0.
Let $T$ be the first time that some job
in ALG is processed for at least $n$ units of work.
Let $G(T)$ be the total flow time plus
energy incurred by ALG up to $T$.
We consider two cases depending on $G(T) \ge k n^3$ or
$G(T) < k n^3$.
If $G(T) \ge k n^3$, Opt reveals
that all jobs are of size $n$. By running at speed 1,
Opt completes all jobs by time $n^2$. The total
flow time plus energy of Opt is at most $n^3 + n^2P(1) \le 2 n^3$,
so ALG is $\Omega(k)$-competitive.

The rest of the proof assumes $G(T) < k n^3$. Let $q_1, q_2, \dots, q_n$
be the amount of work ALG has processed for each of the $n$ jobs.
Without loss of generality, we assume $q_n = n$. Opt reveals that the size of each job $j_i$
is $p_i = q_i + 1$. Thus, at time $T$, ALG has $n$ remaining jobs,
each of size 1. For Opt, it runs at the same speed as ALG during
$[0, T]$ and processes exactly the same job as ALG except on $j_n$.
By distributing the $n$ units of work processed on $j_n$ to
all the $n$ jobs, Opt can complete $j_1, \dots, j_{n-1}$ by time $T$
and the remaining size of $j_n$ is $n$. As Opt is simulating
ALG on all jobs except $j_n$, the total
flow plus energy incurred by Opt up to $T$ is at most $G(T) < kn^3$.

During $[T, T+n^4]$, Opt releases a stream of small jobs.
Specifically, let $\epsilon < \frac{1}{n^5 v^2}$ be any real.
For $i = 1, \dots, \frac{n^4}{\epsilon}$,
a small job $j'_i$ is released at $T + (i-1) \epsilon$ with
size $\epsilon v$. Opt can run at speed $v$ and complete
each small job before the next one is released. Thus, Opt
has at most one small job and $j_n$ remaining at any time during
$[T, T+n^4 ]$.
The flow time plus energy incurred during this period
is $2 n^4 + n^4 P(v)$. Opt can complete $j_n$ by running at
speed 1 during $[T+ n^4, T+n^4+n]$, incurring a cost of $n + n P(1)$.
Thus, the total flow time plus energy of Opt for the whole job sequence is
at most $kn^3 + 2n^4 + n^4P(v) + n + nP(1) = O(n^4 P(v))$.

For ALG, we first show that its total work done on the small jobs
during $[T, T+n^4]$
is at least $n^4v-1$. Otherwise, there are at least $\frac{1}{\epsilon v} > n^5 v$
small jobs not completed by $T+n^4$. The best case is when these jobs
are released during $[T+n^4 - \frac{1}{v}, T+n^4]$ and their total flow time incurred is
$\Omega( n^5 )$. It means that ALG is $\Omega( k )$-competitive
as $n = \ceil{ k P(v)}$.

We call $j_1, \dots, j_n$ \emph{big} jobs and then consider the
number of big jobs completed by ALG by time $T+n^4$.
If ALG completes less that $\frac{1}{2}n + 1$ big jobs by time $T+n^4$,
then ALG has at least $\frac{1}{2}n - 1$ big jobs remaining at any time
during $[T, T+n^4]$. The total flow time of ALG is at
least $\Omega( n^5 )$, meaning that ALG is $\Omega( k )$-competitive.
If ALG completes at least $\frac{1}{2}n + 1$ big jobs by time $T+n^4$,
the total work done by ALG during $[T, T+n^4]$ is at least
$n^4 v -1 + \frac{1}{2}n + 1 $. The total energy used by ALG is at least
\[
P( \frac{ n^4 v + \frac{1}{2}n} {n^4} ) \times n^4 =
P( v + \frac{1}{2n^3} ) \times n^4
\ge P( v + \frac{1}{16 (k P(v))^3} ) \times n^4
\]
The last inequality comes from the fact that
$n = \ceil{k P(v) } \le 2 k P(v)$.
Hence, ALG is at least $\Omega( P( v + \frac{1}{16(kP(v))^3})/P(v) )$-competitive.
\end{proof}

Then, we can apply Lemma~\ref{lem:lower} to obtain the lower bound
for the power function $P(s) = s^\alpha$.

\begin{theorem}
When the power function is $P(s) = s^\alpha$ for some $\alpha >1$,
any algorithm is $\Omega( \alpha^{1/3-\epsilon})$-competitive for
any $0 < \epsilon < 1/3$.
\end{theorem}

\begin{proof}
We apply Lemma~\ref{lem:lower} by putting $k = \alpha^{1/3-\epsilon}$ and
$v = 1$. Then, $P(v) = 1$ and
\[
P(v + \frac{1}{16(kP(v))^3} )/P(v)
= \left( 1 + \frac{1}{16(\alpha^{1/3-\epsilon})^3} \right)^\alpha
= \left( 1 + \frac{1}{16 \alpha^{1-3\epsilon}} \right)^{
  (\alpha^{1-3\epsilon})\times\alpha^{3\epsilon} }
\]
Since $(1+ \frac{1}{16x})^x$ is increasing with $x$
and $\alpha^{1-3\epsilon} \ge 1$, the last term above
is at least $(1+\frac{1}{16})^{\alpha^{3\epsilon}}$.
Thus, $\min\{ k, P(v + \frac{1}{16(kP(v))^3} )/P(v) \}
\ge \min\{ \alpha^{1/3-\epsilon}, (\frac{17}{16})^{\alpha^{3\epsilon}} \}
= \Omega(\alpha^{1/3 -\epsilon})$,
and the theorem follows.
\end{proof}

We also show that for some power function, any algorithm is $\omega(1)$-competitive.
\begin{theorem}
There exists some power function $P$ such that any algorithm
is $\omega(1)$-competitive.
\end{theorem}

\begin{proof}
We want to find a power function $P$ such that for any $k \ge 1$,
there exists a speed $v$ such that $P(v + \frac{1}{16(kP(v))^3})/P(v) \ge k$.
Then by setting
$k$ and $v$ correspondingly to Lemma~\ref{lem:lower}, any algorithm
is at least $k$-competitive for any $k \ge 1$. It implies that
any algorithm is $\omega(1)$-competitive.
For example, consider the power function
\[
P(s) = \frac{1}{ (4(2-s))^{1/4}}\enspace, \enspace\mbox{$0 \le s < 2$}
\]
Let $P'$ be the derivative of $P$.
We can verify that $P'(s) = (P(s))^5$ for all $0 \le s <2$.
For any $k$, let $v \ge 1$ be a speed such that $P(v) \ge 16k^4$.
Then,
\[
P( v + \frac{1}{16(kP(v))^3}) \ge P(v) + P'(v) \frac{1}{16 (k P(v))^3 }
\ge (P(v))^5 \frac{1}{16 (k P(v))^3 } \ge k P(v)
\]
Thus, $P(v + \frac{1}{16(kP(v))^3})/P(v) \ge k$ and the theorem
follows.
\end{proof}

\section{Conclusion}

We show that nonclairvoyant policies can be $O(1)$-competitive in the traditional
power model. However, we showed that in contrast to the case for clairvoyant algorithms,
there are power functions that are sufficiently quickly growing that nonclairvoyant algorithms
can not be $O(1)$-competitive.

One obvious open problem is to reduce the competitive ratio achievable by a nonclairvoyant
algorithm in the case that the cube-root rule holds to something significantly more reasonable
than the rather high bound achieved  here.

The standard/best nonclairvoyant job selection policy
for a fixed speed processor is  Short Elapsed Time First ($\SETF$). The most obvious candidate speed scaling policy would
be to use $\SETF$ for job selection, and 
to run at power somewhat higher than the number of active jobs.
The difficulty with analyzing this speed scaling algorithm is 
that it is hard to find potential functions that interact well
with $\SETF$. It would be interesting to provide an analysis of this algorithm.

\bibliographystyle{plain}
\bibliography{HoLeungCHAN}

\end{document}